\documentclass{aa}

\usepackage[varg]{txfonts}

\include{bookdefs}
\usepackage{graphicx}
\usepackage{xspace}

\renewcommand{\vec}[1]{\mbox{\boldmath$#1$}}

\newcommand{\dif}{\mathrm{d}}

\usepackage{color}
\definecolor{orange}{rgb}{.9,.3,0}
\usepackage{float}

\begin{document}
\title{Fluctuation dynamo amplified by intermittent shear bursts in convectively driven magnetohydrodynamic turbulence}
\author{J. Pratt\inst{1} \and A. Busse\inst{2} \and W.-C. M\"uller\inst{3}}
\institute{Max-Planck-Institut f\"ur Plasmaphysik, 85748 Garching, Germany and Max-Planck-Institut f\"ur Sonnensystemforschung, 37191 Katlenburg-Lindau, Germany \and Faculty of Engineering and the Environment, University of Southampton, UK \and Center for Astronomy and Astrophysics, ER 3-2, TU Berlin, Hardenbergstr. 36, 10623 Berlin, Germany}

\abstract{Intermittent large-scale high-shear flows are found to occur frequently and spontaneously in
{direct numerical} simulations of statistically stationary turbulent Boussinesq magnetohydrodynamic (MHD)
convection.  The energetic steady state of the system is sustained by
{convective driving of the velocity field} and small-scale
dynamo action. {The intermittent emergence of flow structures
 with strong velocity and magnetic shearing generates magnetic} energy at
an elevated rate over time scales longer than the characteristic time
of the large-scale convective motion. {The resilience of magnetic 
energy amplification suggests that intermittent shear bursts  
are a significant driver of dynamo action in turbulent magnetoconvection.}
}

\maketitle

\section{Introduction}

X-ray observations reveal that turbulent convection agitates the outer convection layer of stars  \citep{guedel1997,rnb2007,bohm2008chromospheric,simon2008limits}.  
 Measurements also show that planetary magnetic fields can change in magnitude and orientation \citep{McFadden1995,christensen2009energy,stevenson2010planetary,olson2011complex}.   Dynamo action driven by turbulent convection is accepted as the origin of solar and planetary magnetic fields.  Key physical processes involved in turbulent convection and implicated in the amplification of magnetic fields remain to be identified and practically understood \citep{zeldovich:book,biskamp:book3}.  Helicity, shear, and buoyancy remain intensely interesting to the dynamo problem  \citep{tobias2009,wicht2010,weiss2009solar}.    
 
 Because of the inherent nonlinearity of turbulent plasma flows,
 theoretical explanation of dynamo action is often approached by 
mean-field theory.  
Comparison with three-dimensional numerical simulations verifies and inspires theoretical models \citep{moll_etal:convecdyn,schrinner2005mean,schrinner2007mean,wilkin2007magnetic,harder2005finite,stanley2010dynamo,tcb2011}.
 This work reports on a resilient and newly identified feature of characteristic dynamo action in three-dimensional,
convectively driven, homogeneously turbulent, Boussinesq magnetoconvection
based on pseudospectral direct numerical simulations using the
magnetohydrodynamic (MHD) fluid approximation \citep{chandrasekhar:book}. 

 \section{Simulation} 

Formulation of optimal boundary conditions for simulations of turbulent flows is 
delicate because boundaries strongly influence
the structure and dynamics of the flow. The commonly used
Rayleigh B\'enard boundary conditions impose a temperature gradient to drive turbulent convection by fixing the temperature on impermeable top and
 bottom boundaries.
 For the Reynolds and Rayleigh numbers 
 achievable with current numerical capabilities, the convection-cell
pattern imposed on the flow by Rayleigh B\'enard boundary conditions constrains the development of buoyantly driven turbulence.

The simulation volume considered in this work is confined by quasi-periodic rather than Rayleigh B\'enard boundary conditions.  
   The only deviation from full periodicity in the quasi-periodic boundary conditions is the explicit suppression of mean flows
parallel to gravity, which are removed at each time step.
 Because our simulations are pseudospectral, the mean flow is straightforwardly isolated as the $k=0$ mode in Fourier space, which corresponds to the volume-averaged velocity.
  The aim is to combine the conceptual simplicity of statistical homogeneity with a
 physically natural convective driving of the turbulent flow.   In the flow allowed by the quasi-periodic boundary conditions we identify a process, the shear burst, in our simulation that efficiently amplifies magnetic energy at all spatial scales in convective turbulence.  This process can be relatively subtle, but arises in all cases considered in this work. The simulation model we employ is idealized, but can be viewed as a volume in an astrophysical or geophysical convective turbulent flow that is small in comparison to the pressure scale height.

\begin{figure}
\resizebox{\hsize}{!}{\includegraphics{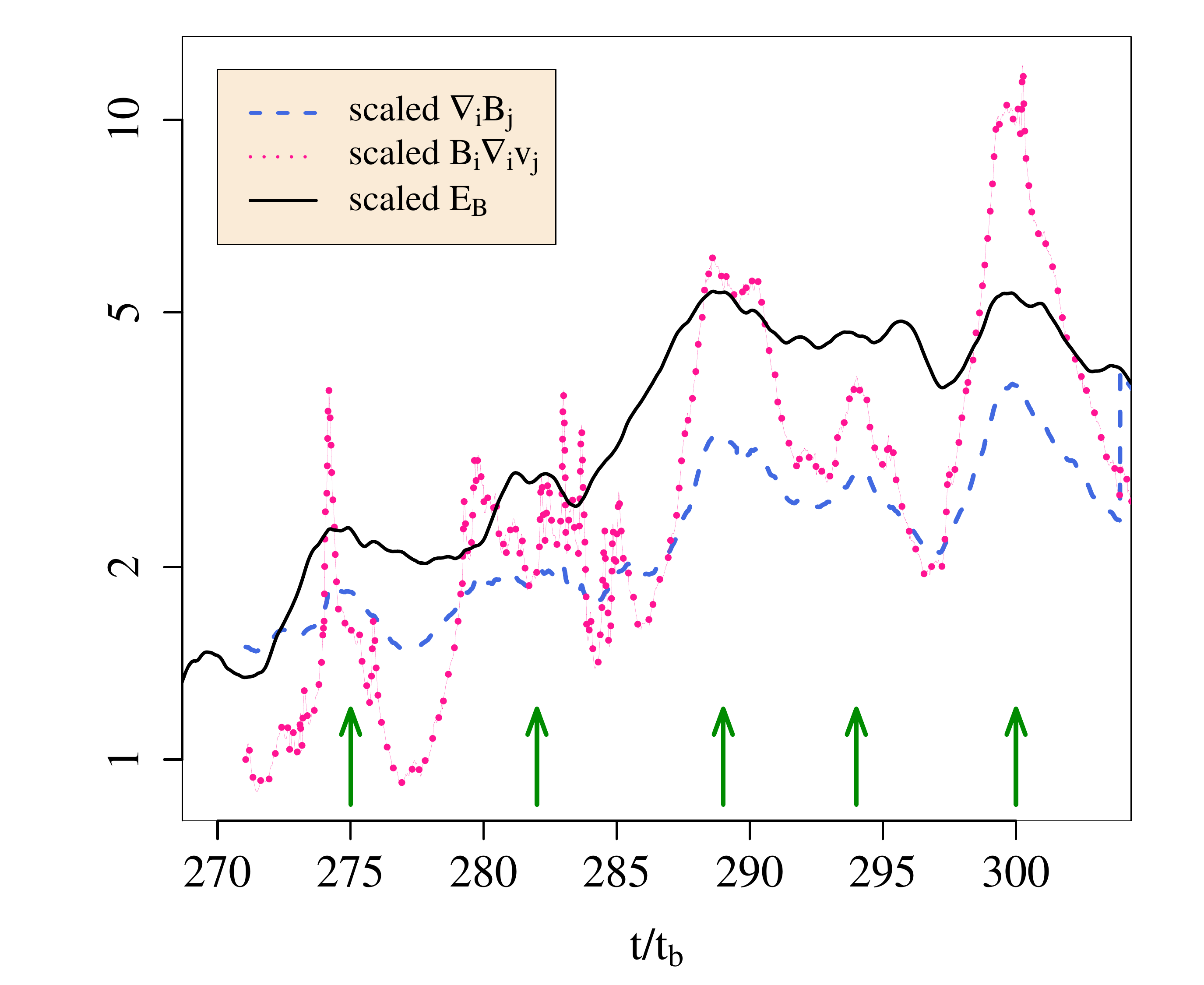}}
\\
\resizebox{\hsize}{!}{\includegraphics{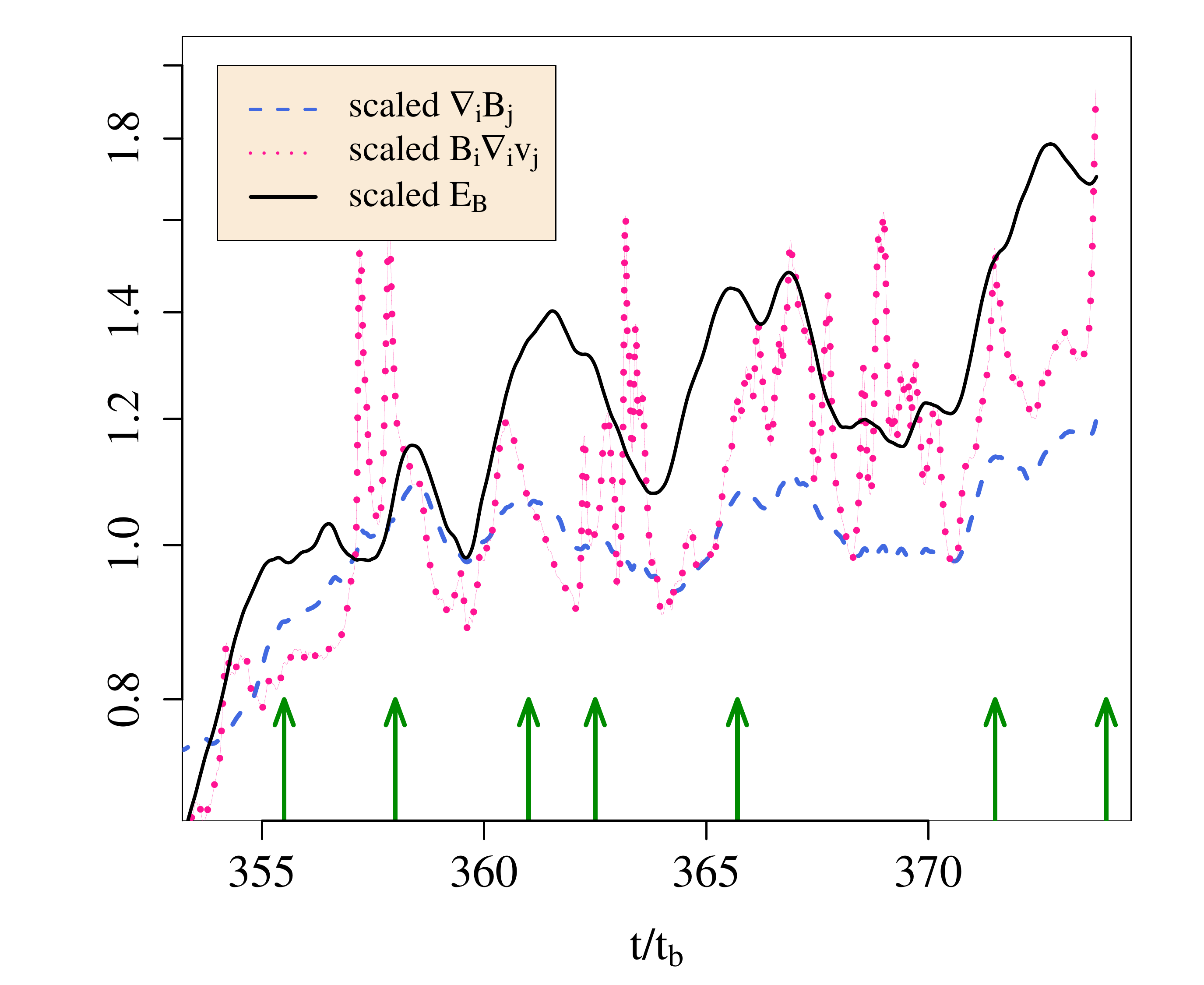}}
\caption{Magnetic energy, $E_{\mathrm B}$, is significantly amplified over long times due to repeated shear bursts in (above) simulation g1 and (below) simulation g11 (parameters described in Table \ref{reynolds_param}). 
The signature of the shear burst, identified here with arrows, is clearly visible in the simultaneous peaks of the magnitude of the magnetic shear tensor $\nabla_i B_j$, and magnetic stretching tensor $B_i \nabla_i u_j$, which have been normalized to their initial-time values for easy comparison.    The magnitude of the velocity shear tensor $\nabla_i
v_j$  closely follows the magnetic shear. The magnitude of these tensors is calculated as a sum of the
squares of the elements.    Shear, and therefore magnetic stretching, drive the intermittent growth
of magnetic energy.
\label{stepgraph}}
\end{figure}

Our system allows the study of a turbulent fluctuation dynamo (also known as a small-scale dynamo) in
detail since the applied boundary conditions permit
shear bursts on large spatial and temporal scales without enforcing a
large-scale structuring of the turbulent flow.  Hundreds of convective
 time scales prove necessary for the study of the shear bursts 
that arise spontaneously in simulations of steady-state
convective MHD turbulence. 
Shear bursts are intermittent and spatially localized around high-shear flows.
 They are driven primarily at multiple large length scales that do not necessarily form a continuous band in wavenumber space, and that vary between bursts.
A single isolated burst is not sufficient to maintain elevated magnetic energy; however, shear bursts can sometimes recur frequently, as shown in Fig. \ref{stepgraph}, providing an elevated
growth of magnetic energy over significant periods of time.
We address the properties of shear bursts and their importance for the understanding of the fluctuation-dynamo
mechanism based on observations from high-resolution direct numerical simulations that
span extended periods of time.

The non-dimensional Boussinesq equations for MHD convection are
\begin{eqnarray} \label{realbmhdc}
\frac{\partial \vec{\omega} }{\partial t} &-& \nabla \times (\vec{v} \times \vec{\omega} +  \vec{j} \times \vec{B}) 
=  \hat{\nu} \nabla^2 \vec{\omega} - \nabla \theta \times \vec{g}_0
\\
\frac{\partial \vec{B} }{\partial t} &-& \nabla \times (\vec{v} \times \vec{B}) =  \hat{\eta} \nabla^2 \vec{B}
\\ \label{thermeq}
\frac{\partial \theta }{\partial t} &+& (\vec{v} \cdot \nabla) \theta = \hat{\kappa} \nabla^2 \theta -  (\vec{v} \cdot \nabla) T_0\\
&& \nabla \cdot \vec{v}=  \nabla \cdot \vec{B}=0 ~~.
\end{eqnarray}
The equations include the solenoidal velocity field $\vec{v}$,
vorticity $\vec{\omega}=\nabla\times\vec{v}$, magnetic field $\vec{B}$, and 
current $\vec{j}=\nabla\times\vec{B}$.  The quantity $\theta$ denotes the temperature fluctuation about
a linear mean temperature profile $T_0(z)$ where $z$ is the direction of
gravity.  In equation \ref{thermeq} this mean temperature provides the convective drive of the system.  In eq. \ref{realbmhdc}, the term including the temperature fluctuation $\theta$ is the buoyancy force.  The vector $\vec{g}_0$ is a unit vector in the direction of gravity.
 These equations are solved using a pseudospectral
method with an adaptive time integration accomplished by a low-storage third-order Runge Kutta
scheme \citep{will80}.

Turbulent convective motion defines the characteristic time and length
scales of the system: the large-scale buoyancy time, $t_{\mathrm b} =(\alpha g
|\nabla T_0|)^{-1/2}$ and temperature gradient length scale
$\mathsf{L}=T_*/\nabla T_0$ where $T_{*}$ is defined as the root-mean-square of
temperature fluctuations $\theta$.  The volume thermal
expansion coefficient at constant pressure is $\alpha$,
 and the gravitational acceleration is $g$ \citep{gibert_etal:T0,wolfprl09}.  The
magnetic field is given in Alfv\'enic units, with Alfv\'en Mach number
$v_0/v_{\mathrm A} = 1$, $v_0=\mathsf{L}/t_{\mathrm b}$.  Three dimensionless parameters
appear in the equations: $\hat{\nu}$, $\hat{\eta}$, and $\hat{\kappa}$.
They derive from the kinematic viscosity $\nu$, the magnetic diffusivity
$\eta$, and thermal diffusivity $\kappa$ and can formally be identified
as the reciprocal value of classical Reynolds number, magnetic Reynolds
number, and P\'eclet numbers, respectively (see Table
\ref{reynolds_param} for definitions). 

To investigate the influence of diffusivities on the shear burst phenomena, parameters 
$\hat{\nu}$, $\hat{\eta}$, and $\hat{\kappa}$ are varied; consequently the simulations
probe values of the Prandtl $\mathsf{Pr}=\hat{\nu}/\hat{\kappa}$ and
 the magnetic Prandtl number $\mathsf{Pr_m}=\hat{\nu}/\hat{\eta}$ spread between 0.5 and 2.  
 The magnetic Prandtl number has been shown to exhibit a significant effect on
small-scale dynamo action \citep{boldyrev_cattaneo:kazantsevdyn,schekbran:magprandtl}; the dependence of the dynamo mechanism
on Prandtl number has been the subject
of several wide-ranging investigations  \citep{schmalzl2002influence,maron2004nonlinear,simitev2005prandtl}.
Realistically small Prandtl numbers cannot be reached with contemporary computer capabilities;  in the solar convection zone expected values are
 $\mathsf{Pr_m} \sim  10^{-4}$ -- $10^{-7}$ and in the earth's core
 $\mathsf{Pr_m} \sim 10^{-6} $.  Simultaneously, Reynolds numbers are generally expected to be larger than can be computationally reached:  $\mathsf{Re} \sim 10^{13}$ in the solar convection zone and  $\mathsf{Re} \sim 10^{8}$ in the earth's core \citep[summarized in][]{angeladiss09}.   Because of this discrepancy,
 the dynamical ranges of fluctuations in modern simulations are smaller than those expected in real systems.  Our
simulations thus present a first impression of the role of shear bursts
for astrophysical dynamos.  
\begin{figure}
\resizebox{\hsize}{!}{\includegraphics{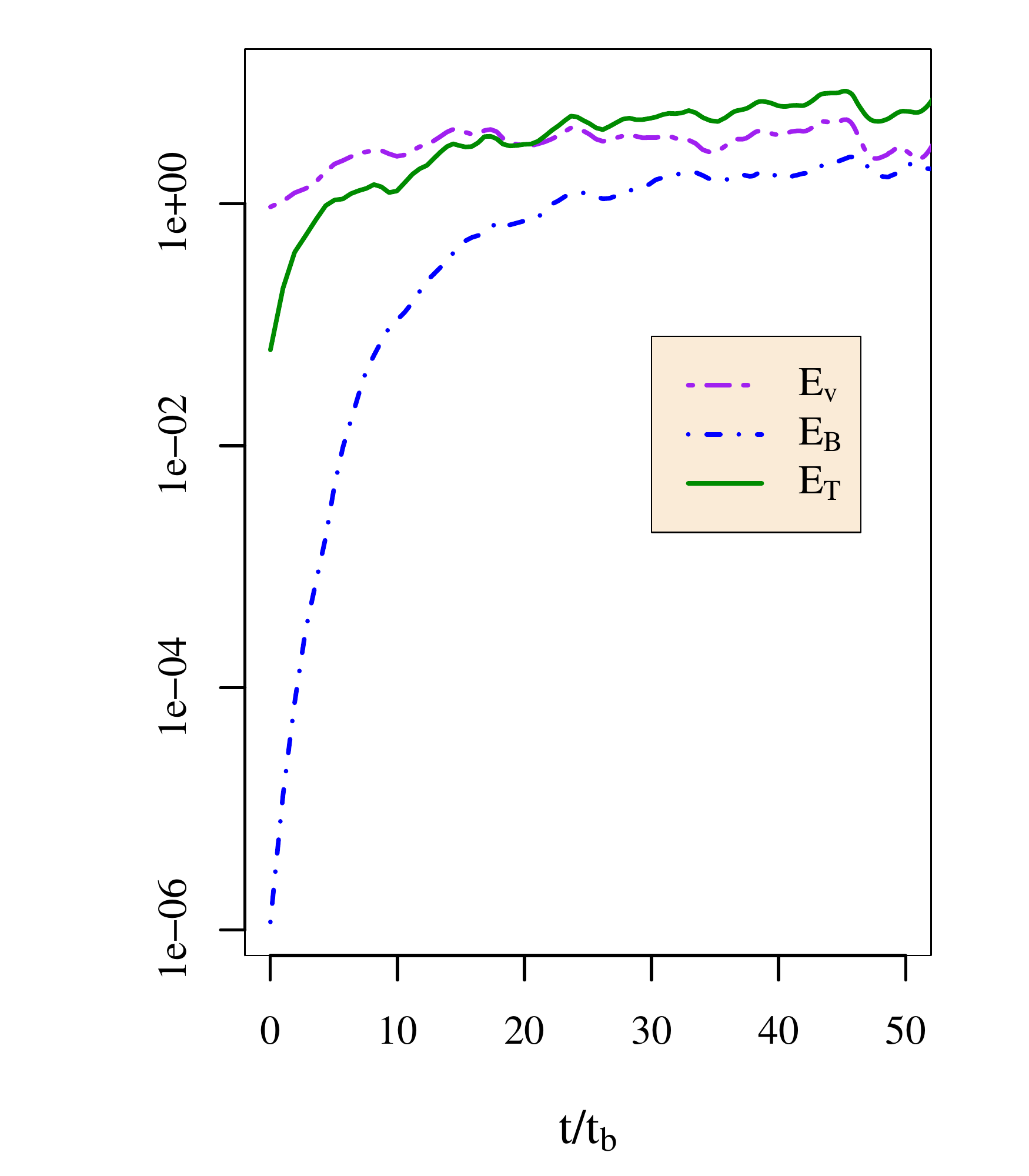}}
\caption{
Total kinetic, magnetic and thermal energies during the initial growth
phase of the dynamo in simulation g2.
\label{en_rise} }
\end{figure} 

\section{Results\label{resultsection}}

The numerical turbulence data in this work results from a set of
simulations conducted with grid size $512^3$, which constitutes high resolution for the extremely long times treated here. These simulations are
performed in a quasi-periodic slab or cube; the cube has a side of $2\pi$, and the slab has slightly larger $x$- and $y$-directions of $(2\pi)^{3/2}$  to allow for
the well-defined initial onset of the convective instability driving the
turbulence \citep{chandrasekhar:book}.  Our boundary conditions inhibit
the formation of viscous boundary layers, which appear when impermeable
boundary conditions are employed.  The dissipative coefficients
$\hat{\nu}$, $\hat{\eta}$, and $\hat{\kappa}$ parametrize the extent
of the turbulent inertial range of scales, and in each simulation are chosen to be as small 
as possible so that the resolution constraint of $k_\mathrm{max}\eta_{\mathsf{kol}}>1.5$ is still satisfied \citep{pope:book}.  Here,
$k_\mathrm{max}$ is the largest resolvable wavenumber allowed by
numerical resolution and $\eta_{\mathsf{kol}}$ is
the Kolmogorov length scale. 
 We performed simulations for the wide range of parameters summarized in Table \ref{reynolds_param} in order 
to test for a possible dependence of the shear burst phenomenon on the 
Reynolds numbers and Prandtl numbers.  Shear bursts occurred in all of the simulations listed.
The Rayleigh number,
characterizing the dynamical importance of buoyancy in Rayleigh-B\`enard
configurations, is of limited informative value for the present
quasi-periodic system.

\begin{table*}
\begin{center}
\caption{Dimensionless simulation parameters \label{reynolds_param}}
\begin{tabular}{lccccccccccc}
\hline\hline
Simulation                                                          & g1  & g2 &  g3 &  g4 & g5   & g6   & g7   & g8   & g9 & g10 & g11 
\\ \hline
$\mathsf{Re}$ ($\times 10^3$)           &  3.2 & 5.4 & 5.2  & 5.6 &  2.3    & 5.1 & 4.0 & 2.4 & 1.3 & 6.1 & 2.4
\\
\hline
$\mathsf{Re_m}=\mathsf{Pr_mRe}$ ($\times 10^3$)                     & 6.4  & 5.4 & 5.2  & 2.8 &  4.0  & 7.7 & 8.0 & 4.2 & 3.9 & 9.8 & 4.8
\\ \hline
$\mathsf{Pe}=\mathsf{PrRe}$ ($\times 10^3$)                         & 3.2  & 2.7& 5.2 & 5.6 &  4.0   & 10.2 & 6.0 & 4.2 & 1.3 & 9.8 & 3.1
\\ \hline
$\mathsf{Pr=\hat{\nu}/\hat{\kappa}}$                                & 1    & 0.5   & 1   & 1   & 1.73  & 2 & 1.5 &1.76 & 1 & 1.6 &1.3
\\ \hline
$\mathsf{Pr_m=\hat{\nu}/\hat{\eta}}$                                & 2    & 1     & 1   & 0.5 & 1.73  & 1.5 & 2 &1.76 & 3 & 1.6 & 2
\\ \hline
$\mathsf{Ra}=(\hat{\nu}\hat{\kappa})^{-1}$  ($\times 10^{5}$)       & 2.5  & 2.2   & 2.5 & 4.4  & 1.4  & 2.2 &1.7 & 0.9 & 0.3 & 3.8 & 1.7
\\ \hline
 $k_{\mathsf{max}} \eta_{\mathsf{kol}}  $                                          & 2.0  & 1.6   & 1.8 & 1.7 & 2.1   &  2.5 & 2.4 & 3.2 & 4.1 & 1.7 & 2.0
\\ \hline
\end{tabular}
\end{center}
\tablefoot{Dimensionless simulation parameters include the magnetic
Reynolds number $\mathsf{Re_m}$, P\'eclet number $\mathsf{Pe}$, Prandtl number
$\mathsf{Pr}$, magnetic Prandtl number $\mathsf{Pr_m}$, and Rayleigh number $\mathsf{Ra}$. The
Reynolds number is defined as $\mathsf{Re= \langle E_v^{1/2} L\rangle /\nu} $, where $\mathsf{L}$ is the instantaneous temperature gradient length scale and the brackets indicate a time-average.}
\end{table*}
The initial state of the simulations contains 
fluctuations in a number of small $k$ modes for the velocity,
magnetic field, and temperature.  The initial ratio of kinetic to
magnetic energy of turbulent fluctuations is $10^{6}$ with the kinetic
energy of order unity. Fig. \ref{en_rise} shows a typical example of
the initial time-evolution of kinetic energy
$E_{\mathsf v}=V^{-1}\int_V\dif V \mathsf{v}^2/2$, magnetic energy $E_{B}=V^{-1}\int_V\dif V B^2$,
and thermal energy $E_{T}=V^{-1}\int_V\dif V\theta^2$ taken from simulation g2. The thermal energy should be understood
as the variance of temperature fluctuations. Magnetic
energy rises quickly due to small-scale dynamo action and saturates at
$E_{\mathsf v}/E_{B}\approx O(1)$, characteristic of the quasi-stationary turbulent
state of the MHD flow.
  
In Fig. \ref{en_rise} the global energies of the steady-state system
evolve with fluctuations due to the convective motion.  After a simulation reaches steady state, energies fluctuate on the order of 10\%,
 with a period of a 1-2 buoyancy times \citep[see also Figure 5 of][]{cat99}.
During steady-state turbulent convection, we begin to observe a pattern of
spontaneous longer periods (5-20 $t_{\mathrm{b}}$) of significant growth
in the global magnetic, thermal and kinetic energies.  During the kinematic stage of the dynamo, before steady-state convection is
reached, we find no evidence of these physically interesting periods of energy growth.  The net energy
variation during one such period can differ greatly, but we observe the energy to reach 10 times
the steady-state energy level during particularly strong instances.  For example, these periods of unusually elevated
energy growth occur 15 times, unevenly spaced over a time span of
225 $t_{\mathrm{b}}$, in the simulation g1.  We associate the growth of energy during these periods with the shear burst phenomena.  Shear bursts can occur in close sequence, but do not universally do so.  The system can be regarded as statistically steady over periods of time significantly longer than the duration of a shear burst.  

\begin{figure}
\resizebox{\hsize}{!}{\includegraphics{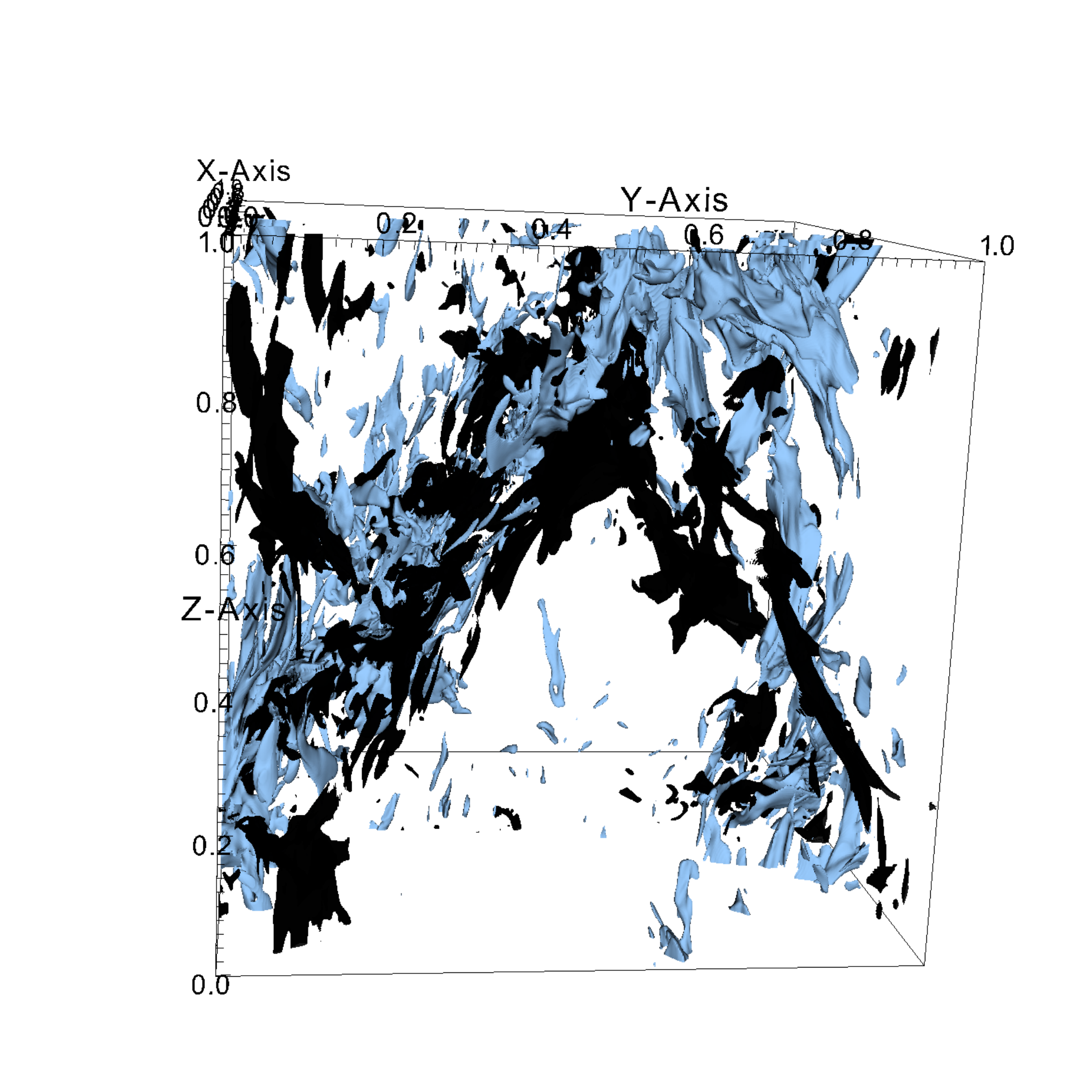}}
\caption{A hot (black) flow moves vigorously along a path upward from the bottom left corner, bending through the middle of the volume, and eventually turning downward again, shearing against a cold (light blue) flow.  The shear along these opposing flow structures drives energy production during a shear burst.
\label{t311shear}}
\end{figure}

A shear burst centers around a period of increased growth of magnetic energy that is accompanied by growth of both magnetic
shear and magnetic stretching.  The growth rate of magnetic energy during a shear burst is uneven, and can vary wildly between shear bursts in the same simulation.  Preceding the growth of magnetic energy, a coherent flow
structure forms that has the appearance of high-velocity, hot or cold coherent streams, in contrast to the typical situation with many
smoothly convecting plumes of hot and cold fluid.  These high-energy streams become strongly aligned
in space, producing regions of high and increasing velocity and magnetic
shear.   The nonlinear shape and orientation of the high-energy streams differ for each shear burst, displaying no preferred direction.  
 The coherent flows that form in one instance of a shear burst are depicted in Fig. \ref{t311shear}.

\begin{figure}
\resizebox{\hsize}{!}{\includegraphics{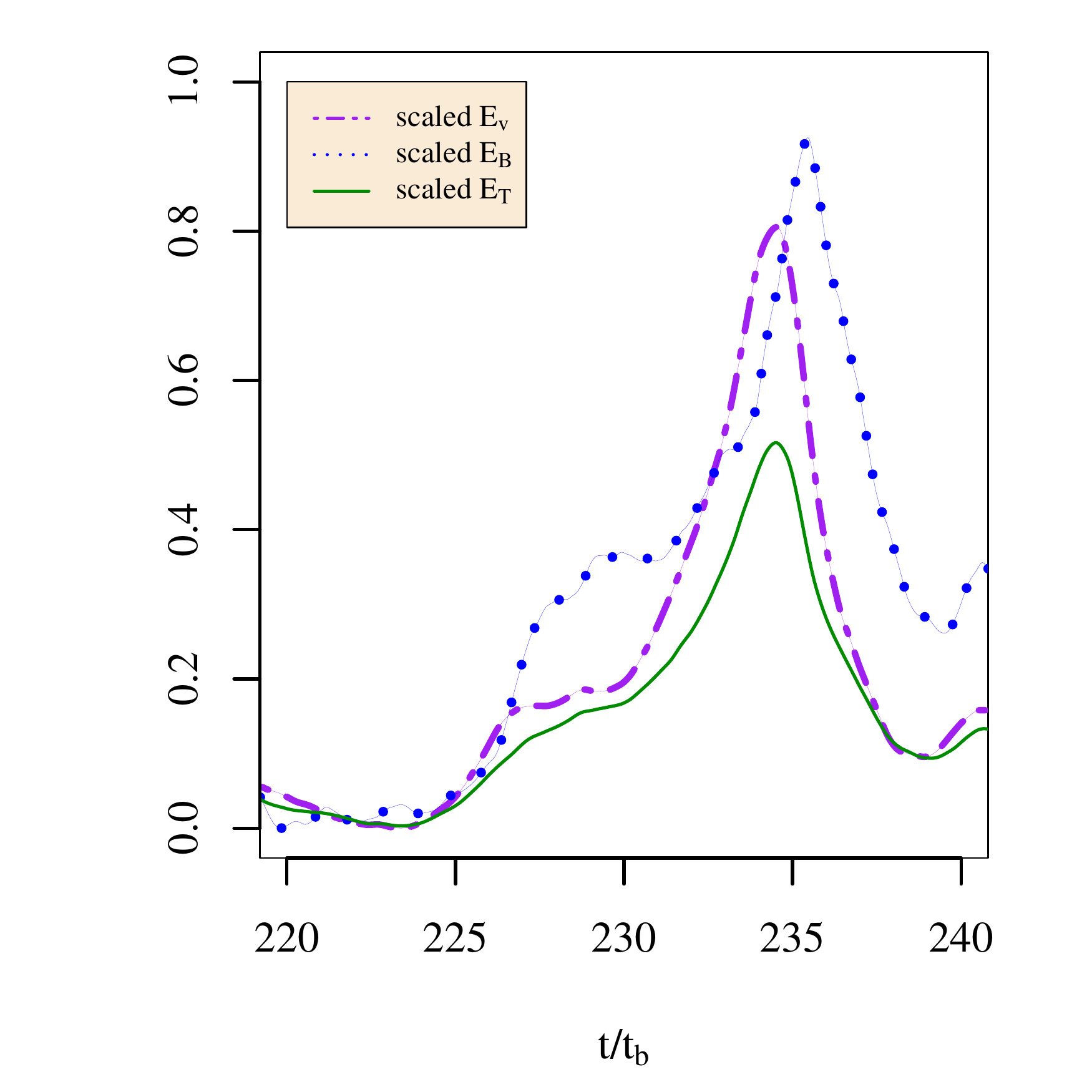}}
\caption{The global kinetic energy $E_{\mathrm{v}}$, magnetic
energy $E_{\mathrm{B}}$, and thermal energy $E_{\mathrm{T}}$ scaled to run between 0 and 1, for the
span of a typical shear burst in simulation g8.  
 \label{en_chop1}}
\end{figure}
Shear causes magnetic field-line stretching, and thus amplification of magnetic energy \citep{gilbert1995stretch,cattaneo1999origin}. In Fig. \ref{stepgraph} each shear burst can be defined
by a peak in magnetic shear that correlates with an increase in magnetic energy.
Fig. \ref{en_chop1} allows for closer inspection of 
 the increase of magnetic, kinetic, and thermal
 energies for a typical isolated shear burst in simulation g8; between $t=225$ and $t=235$ the energies increase by a factor of three.   Individual shear bursts can last from a couple buoyancy times to a couple tens of buoyancy times.  At the peak of magnetic and kinetic energies, high
energy hot and cold shearing is at its most vigorous.  The flow structures lose their alignment, slow down, and ultimately break-up.    The peak of global energy in
Fig. \ref{en_chop1} represents the beginning of the break-up of flow structures.  The
 break-up of the fast streams spurs a slow decline in global
energies.   After the shear burst has dissipated, the energies dissipate until the steady-state level maintained by the fluctuation dynamo has been reached.  Shear bursts can overlap in time, and also can occur closely in sequence, as shown in Fig. \ref{stepgraph}.   
 
 The lifetime and magnitude of
the energy growth, in particular the peak in thermal energy, can vary greatly between simulations and even between shear bursts in the same simulation.
This shows no apparent dependence on the Prandtl numbers.  That the Prandtl numbers do not directly impact the shear burst phenomena is surprising because the Prandtl numbers express the ratio of turbulent intensities and dynamic ranges of the respective turbulent fields.  This relationship between Prandtl numbers and the turbulence can be understood by relating the Prandtl numbers to the ratios of Reynolds numbers, $\mathsf{Pr}=\mathsf{Pe}/\mathsf{Re}$ and $\mathsf{Pr_m}=\mathsf{Re_m}/\mathsf{Re}$, where the P\'eclet number can be regarded as the same structure as a Reynolds number for thermal fluctuations.

The characteristic length-scale of Boussinesq convection is the Bolgiano-Obukhov length, {$\ell_{\mathrm{bo}}=\mathsf{\epsilon_{\mathrm{v}}^{5/4}/\epsilon_{\mathrm{T}}^{3/4}}$} that  separates convectively-driven scales of the flow {$\ell > \ell_{\mathrm{bo}}$} from the range of scales where the temperature fluctuations behave as a passive scalar $\ell < \ell_{\mathrm{bo}}$.  In this definition $\mathsf{\epsilon_{\mathrm{v}}}$ and $\mathsf{\epsilon_{\mathrm{ T}}}$ are the kinetic and thermal energy dissipations respectively.    Typically in our convection simulations $\ell_{\mathrm{bo}}$ is comparable to the system size so only the largest scales in the flow are convectively driven. The shear bursts are large-scale phenomena by this classification, but are not dominated by the convective motion.  This suggests an explanation for the apparent insensitivity to changes in the Prandtl numbers.   

\begin{figure}
\resizebox{\hsize}{!}{\includegraphics{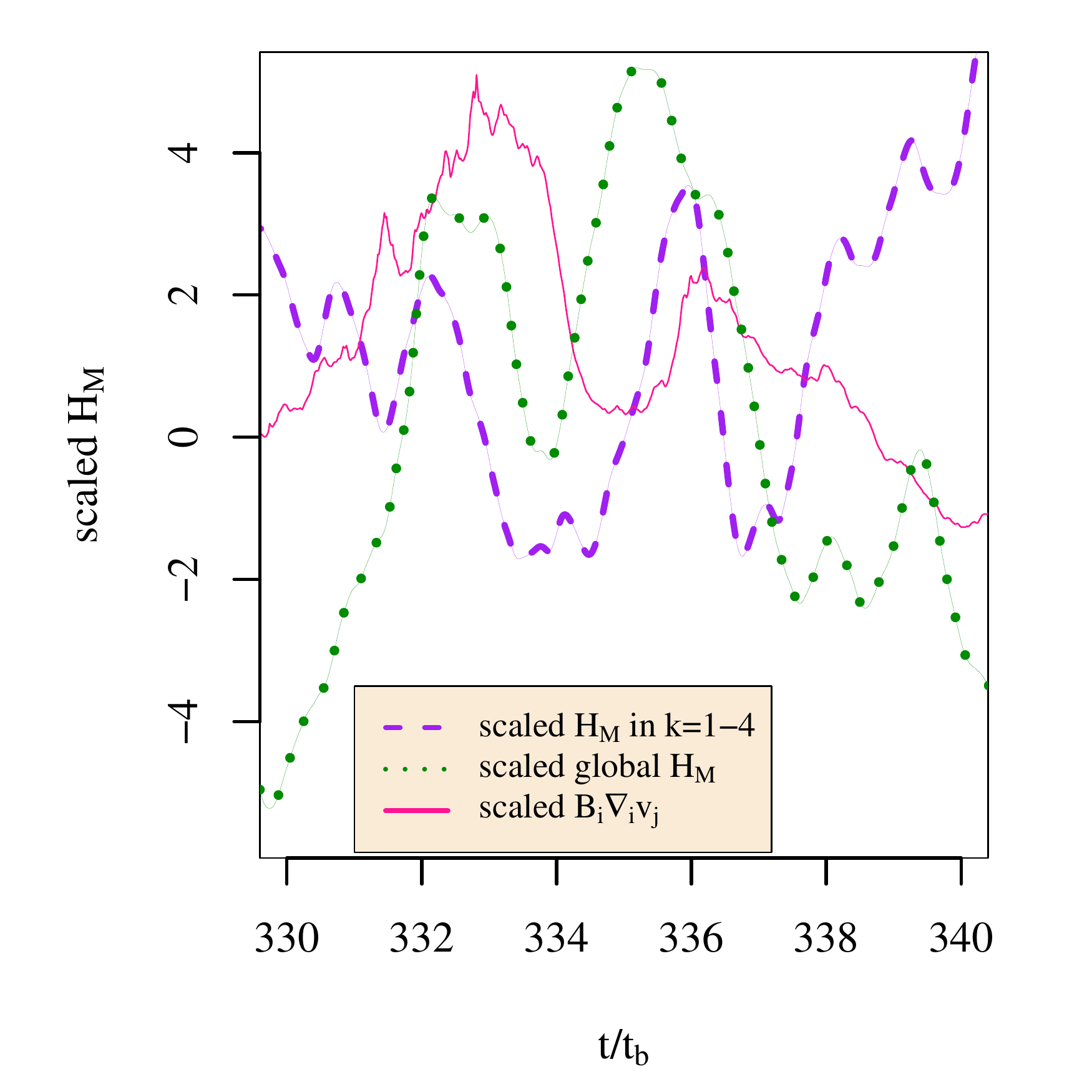}}
\caption{The characteristic double peak from two closely spaced shear bursts in simulation g11 is visible in magnetic helicity.
The magnetic helicity $H_{\mathrm{M}}$ at low-wavenumbers $k$ is shown scaled to its initial time-value with sign preserved. 
Magnetic helicity quantities are shown scaled to their initial time values in a way that preserves sign. The magnitude of the magnetic stretching tensor
 is shown shifted and scaled to fit the scale of the magnetic helicity, to provide a time-reference for the shear bursts.  
 \label{hbspectra_glu3}}
\end{figure}

Although insensitive to the Prandtl numbers, the shear bursts interact nonlinearly with the turbulent environment, mainly via large-scale magnetic structures.  This is reflected in the behavior of magnetic helicity, $H_M=V^{-1}\int_V\dif V \vec{A} \cdot \vec{B}$, which measures the linkage and knottiness of the magnetic field-lines \citep{biskamp:book2,moffatbook}.
A signature of the shear burst is the growth of global magnetic helicity as the shear flows strengthen.
 Magnetic helicity is not conserved in the dissipative system we study,
and this growth of magnetic helicity typically exceeds more than a
standard deviation from the average magnetic helicity over the
time-span of the simulation.  A peak of global magnetic helicity frequently shortly precedes or coincides with a
shear burst.  Fig. \ref{hbspectra_glu3} shows the typical
time-evolution of the magnetic stretching against the growth of global magnetic helicity and magnetic
helicity at the largest scales.  In the time pictured, two shear-busts occur within 5 $t_{\mathrm b}$, and a clear double-peak structure is also visible in the magnetic helicity.

\begin{figure}
\resizebox{\hsize}{!}{\includegraphics{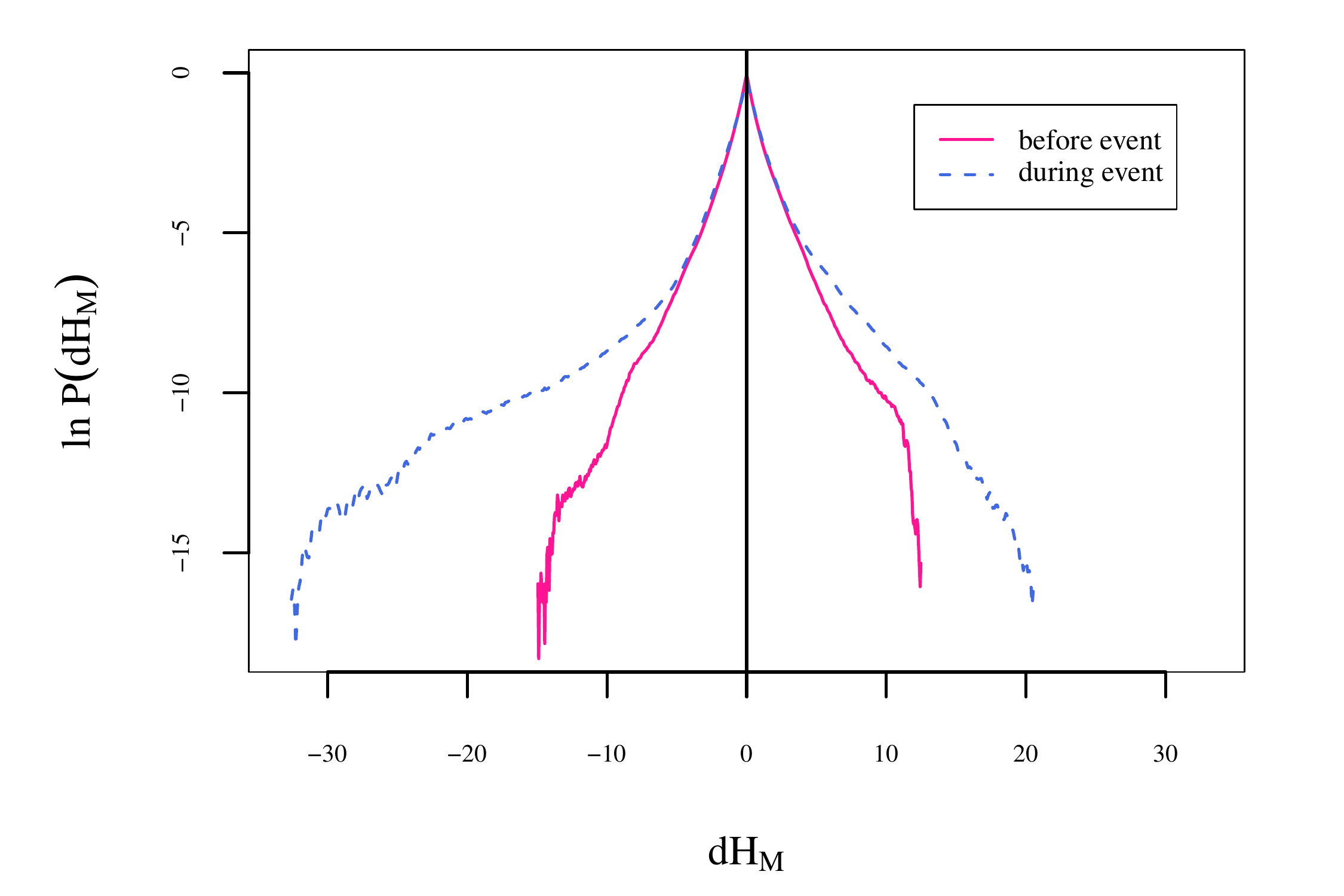}}
\\
\resizebox{\hsize}{!}{\includegraphics{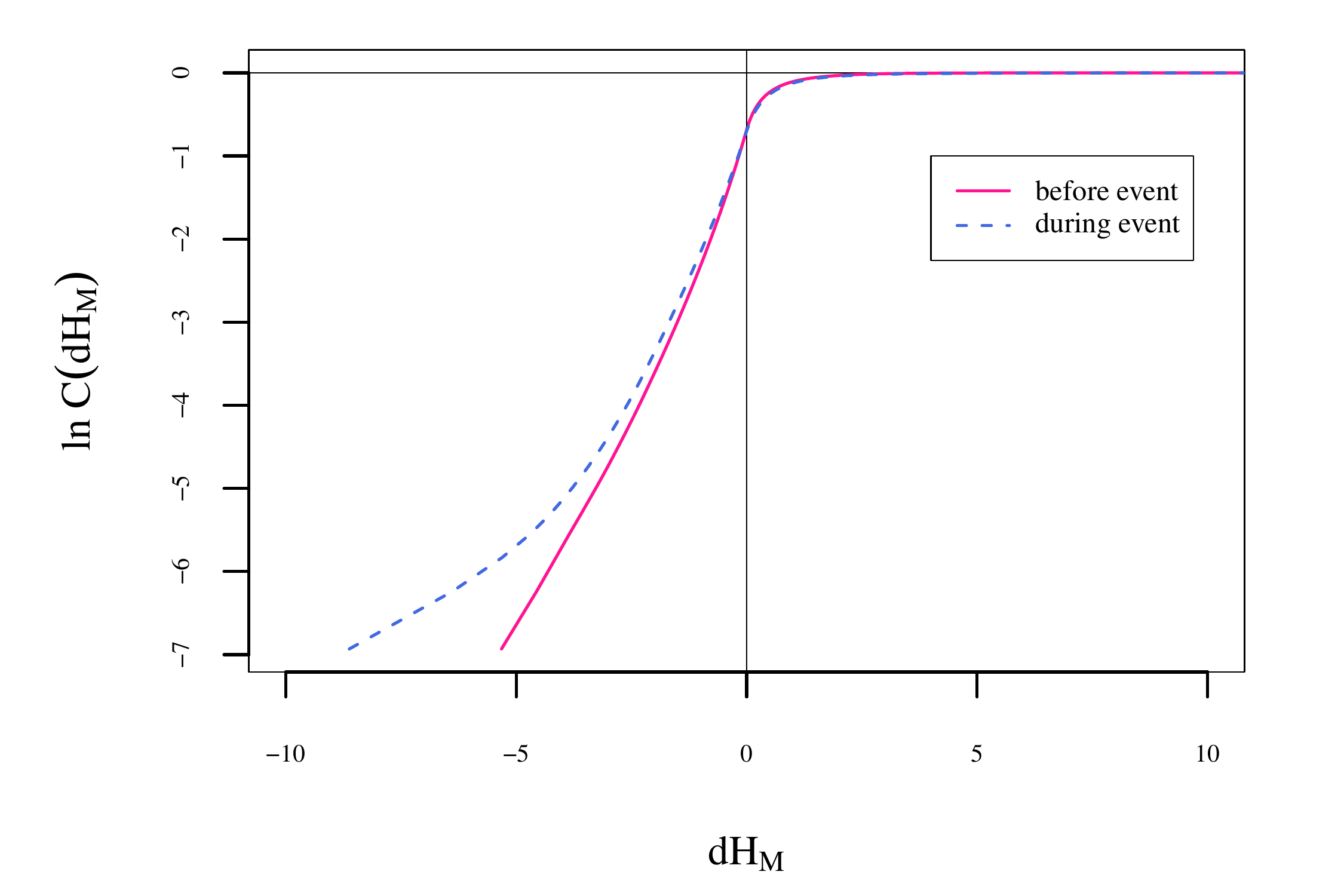}}
\caption{(Above) probability distribution  and (below) cumulative
distribution of local magnetic helicity, \emph{i.e.}
$dH_{\mathrm{M}}=\vec{A} \cdot \vec{B}$ in the simulation volume before
and during a shear burst event in simulation g1.  During this shear burst, as the global magnetic helicity grows, the tail of negative magnetic helicity grows,
and grows much higher than the positive tail. 
\label{mabpdfs}}
\end{figure}

The magnetic helicity grows particularly in low wavenumbers
$k$, associated with the growth of an isolated
structure with a strong helicity polarity; this low-$k$ growth is a
signature of an ongoing inverse spectral transfer of magnetic helicity common for 3D MHD systems \citep{muller2012inverse,biskamp:book3,alexakis2008inverse}.
  The dramatic change in the bias of magnetic helicity in the system during
one shear burst is shown in Fig. \ref{mabpdfs}; in Fig. \ref{hbspectra_glu3}
the large-scale magnetic helicity of the structures spawned also has negative polarity for the two shear bursts pictured.
  Large-scale
magnetic helicity structures persist longer than the high-energy shear streams, and longer than it takes for the global energies to taper off.  Because the magnetic helicity experiences an inverse cascade and our system has small dissipation this is theoretically expected.

\begin{figure}
\resizebox{\hsize}{!}{\includegraphics{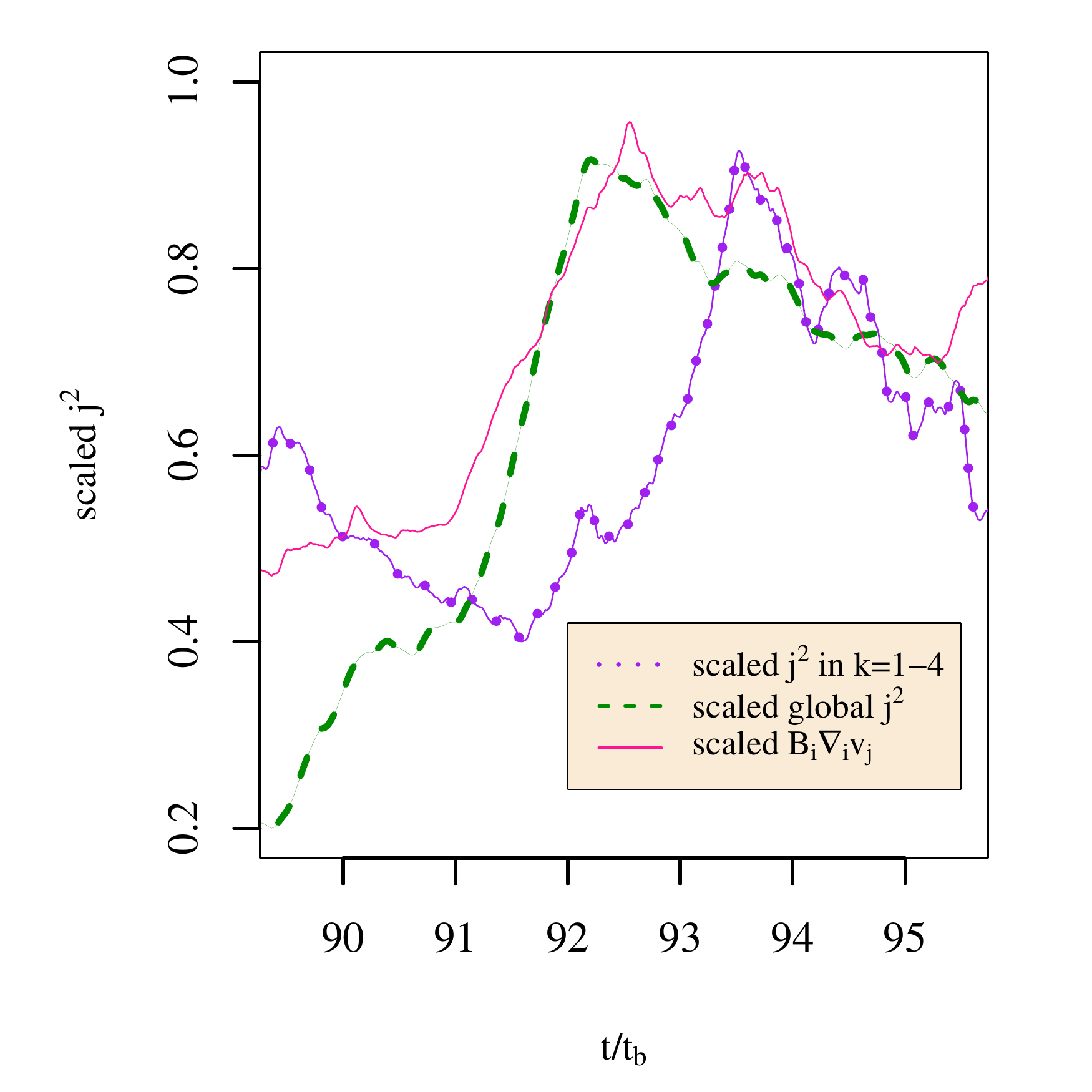}}
\caption{Current-squared in the lowest wavenumbers $k=1,2,3,4$ scaled to its initial time-value during a shear burst in simulation g2.  
Global current-squared and magnetic shear are shown scaled and shifted to fit on the same scale 
for reference. \label{curspectra_g2}}
\end{figure}

Shear bursts generate significant currents through magnetic shear, which change the global magnetic helicity.
When no shear burst is present in the system small filaments of high current are common and likely indicate slow reconnection
on small scales.  However when a shear burst grows, large-scale high-current structures 
 grow at the same time.  A typical growth in current around a shear burst is shown in Fig. \ref{curspectra_g2}.

 In simulations similar to those discussed in this work, but performed with fully-periodic boundary conditions (sometimes called
  homogenous Rayleigh B\'enard boundary conditions), the 
macroscopic elevator instability as it is presented by \citet{calzavarini_etal:elevator2,calzavarini_etal:elevator,danthesis} can be readily identified. 
The elevator instability is an exact solution to the equations of motion in the homogeneous system.
It is an extreme realization of the fact that the
homogeneous, incompressible flow can gain vast amounts of energy by coherent
large-scale vertical motions.  The elevator instability creates parallel, vertical jets ($k_z=0$) of alternating direction throughout the volume that significantly degrade the quality of turbulence statistics.   The flow pattern created by this elevator instability fully destroys
the original, natural flow field.  The instability can be eliminated in homogeneous Boussinesq systems
 by considering a horizontal mean temperature gradient \citep{wolfprl09}.

In the quasi-periodic simulations presented in this work, we find no evidence of this instability although we follow the simulations for extremely long times.  
Mean flows parallel to gravity are manually suppressed in our quasi-periodic set-up.
In contrast to the elevator instability, shear bursts are embedded into the turbulence
and have limited coherence and lifetime with regard to the full flow-field.
Shear bursts do not exhibit exponential growth of energy nor is their growth rate dependent on clear system parameters like the Prandtl number.  The shear burst is superficially similar to the elevator instability because both involve coherent flows.  However, the coherent flows associated with a shear burst do not display a preference for any fixed spatial direction, but follow a sometimes complicated, curved path in a localized section of the simulation volume.  This flow path can change during the evolution of the shear burst, and is different for each shear burst.   When the violent elevator instability is present, it is likely to mask finer-scale processes like the shear burst.

\section{Conclusions}

We have isolated a basic mechanism of dynamo action in
MHD convection that operates through spontaneously developing,
intermittent bursts of high shear during steady-state dynamo action.  Because this process occurs in
all simulations considered here, it is of potential importance for astrophysical small-scale
dynamo action in turbulent convection scenarios.  Shear bursts consist of the
formation of coherent, highly-sheared flows along-side magnetic
structures with a strong magnetic-helicity polarity bias.  The slow growth and eventual decay of the magnetic helicity structure is
key to the shear burst phenomenon.  The increasing shear
causes a gradual rise in energy on all spatial scales due to magnetic stretching in the system
over several buoyancy times.  After some time, the shear flows lose their alignment and decay.  Once the shear flows are destroyed, the elevated energy dissipates over
several buoyancy times.  Closely spaced shear bursts can occur in a series, creating
long periods of time where magnetic energy is elevated significantly above the steady state.

\vspace{7mm}

\begin{acknowledgements}
{This work has been supported by the Max-Planck
Society in the framework of the Inter-institutional
Research Initiative Turbulent transport and ion
heating, reconnection and electron acceleration in
solar and fusion plasmas of the MPI for Solar System
Research, Katlenburg-Lindau, and the Institute
for Plasma Physics, Garching (project MIFIF-
A-AERO8047).  Simulations were performed on the VIP computer system at the Rechenzentrum Garching of the Max Planck Society. }
\end{acknowledgements}
\bibliographystyle{aa}

\end{document}